# Employing Agent Beliefs during Fault Diagnosis for IEC 61499 Industrial Cyber-Physical Systems

Barry Dowdeswell[1], Roopak Sinha[1], Dennis Jarvis[2], Jacqueline Jarvis[2], Stephen G. MacDonell[1]

[1]School of Engineering, Computing and Mathematical Sciences,
Auckland University of Technology, Auckland, New Zealand.
barry.dowdeswell@aut.ac.nz, roopak.sinha@aut.ac.nz, stephen.macdonell@aut.ac.nz.
[2]School of Engineering and Technology, Central Queensland University Brisbane, Australia.
d.jarvis@cqu.edu.au, j.jarvis@cqu.edu.au.

**Abstract**

*We have come to rely on industrial-scale cyber-physical systems more and more to manage tasks and machinery in safety-critical situations. Efficient, reliable fault identification and management has become a critical factor in the design of these increasingly sophisticated and complex devices. Teams of co-operating software agents are one way to coordinate the flow of diagnostic information gathered during fault-finding. By wielding domain knowledge of the software architecture used to construct the system, agents build and refine their beliefs about the location and root cause of faults. This paper examines how agents constructed within the GORITE Multi-Agent Framework create and refine their beliefs. We demonstrate three different belief structures implemented within our Fault Diagnostic Engine, showing how each supports a distinct aspect of the agent's reasoning. Using domain knowledge of the IEC 61499 Function Block architecture, agents are able to examine and rigorously evaluate both individual components and entire sub-systems.*

## 1. INTRODUCTION

Industrial-scale Cyber-Physical Systems (ICPS) rely on sensors and actuators that interact with computing elements to manage complex tasks. When multiple ICPS communicate and work together, they are able to facilitate manufacturing and control operations that far exceed the capabilities of earlier embedded machinery controllers. However, ICPS demonstrate a level of complexity that demands sophisticated fault identification and diagnosis to prevent catastrophic failures [1], [2].

Multi-Agent Systems (MAS) are computers that employ software entities known as *agents* [3]. Agents working together within a Fault Diagnostic Engine (FDE) demonstrate capabilities that make them ideal for carrying out fault monitoring, identification and diagnosis for ICPS. Wooldridge [4] describes two characteristics of agents that are important for this task. Firstly, agents situated in an environment are capable of semi-autonomous behavior. This allows them to follow fault evidence trails by themselves through a malfunctioning ICPS. By observing and evaluating fault symptoms without assistance from humans, they attempt to reconcile the way the ICPS is operating against known profiles of acceptable performance. Secondly, by co-ordinating their actions with other agents operating within the same environment, they can co-operatively share both the analysis tasks and evidence gathering to ultimately present a diagnosis of the faults.

Central to these capabilities is the way that an agent is able to form and manage *beliefs* about what it is observing. Beliefs are opinions held by an agent, formed during a methodical examination and testing of individual parts of the ICPS they are investigating. Beliefs are dynamic [5]. Previously-held beliefs may be reinforced or discarded, based on new evidence that an agent has captured. This progressive refinement adds both integrity and weight to beliefs, allowing the relative probability of alternate fault hypotheses to be considered in the light of the opinions the agent now holds.

This paper presents the design and application of the belief and reasoning structures used by the software agents in our FDE. We profile three different belief structures that agents employ to interact with a Function Block Application (FBA) constructed from IEC 61499 Function Blocks (FBs) [6]. We demonstrate how the beliefs the agents hold enable them to navigate models they have created that capture how the FBA is constructed from individual FBs. Agents can discern how the component parts of the application are interconnected as well as how to exercise components while checking for faults. Coupled with a separate belief model that defines the skills needed to interact with the FBs during diagnosis, the agents also use a third belief structure to organize their findings. By considering the evidence captured in these complementary belief structures, agents can propose which faults have occurred with higher confidence. This technique illustrates how agents can dynamically monitor systems, updating their beliefs and reconfiguring the control layer from the execution layer. This represents a significant enhancement of the interaction scheme employed in [7], [8] where execution layer agents only initiated function block controlled actions.

Section 2 introduces the characteristics of IEC 61499 Function Blocks and how ICPS using them are crafted. A Heating, Ventilation and Air-Conditioning (HVAC) system constructed from FBs is used to illustrate the types of faults that can occur. Section 3 introduces Multi-Agent Systems, profiling the way our agents use Belief-Desire-Intention (BDI) paradigms to orchestrate their behavior. The three belief structures are explained and contrasted, showing how agents use them to interact with FBAs during diagnostic explorations. Section 4 presents our conclusions and future research directions.

## 2. ICPS BUILT WITH IEC 61499 FUNCTION BLOCKS

ICPS constructed with an IEC 61499 software architecture co-ordinate tasks within their physical environment using a range of sophisticated sensors and actuators. FBs are object-oriented software entities, designed to implement the control logic required to interface with individual sensors. They also co-ordinate the movements of electromechanical actuators such as motors and ducting vents to perform work in their environment [9], [10].

Typical HVAC installations rely on multiple self-managing controllers, deployed in different parts of a building. Fig. 1 illustrates the HVAC sub-systems which provide heating and cooling to a building. Each room controller communicates environmental information to a centralized controller that is responsible for delivering enough warm or cool air to each zone to meet the needs of the occupants. Achieving the desired temperature and humidity requires delicate control of air flows. The HVAC Central Controller typically relies on Proportional Control (PI) algorithms [11] driven by telemetry captured from temperature sensors in each zone.

Fig. 2 details the FBs used to construct a single room controller. The Z_CONTROLLER FB is constructed from a template called a *Basic Function Block* (BFB). The IEC 61499 design convention is to document *Input Events* and *Data Inputs* on the left of the FB symbol. The data connections between individual FBs are shown in blue while event triggers are shown in red. Dotted lines indicate connections to another sub-application that runs the HVAC Main controller on a different computer. The input event CMD_UP receives an event from Z_SWITCHES to notify it that a room occupant has asked for the temperature to be increased. Similarly, when the temperature sensor monitored by Z_TEMPERATURE reports a temperature change, Z_CONTROLLER is notified by an event received by TEMP_CHANGED. The value of the new temperature is available on the data input TEMP as an IEEE-format Real number. Similarly, *Output Events* and *Data*

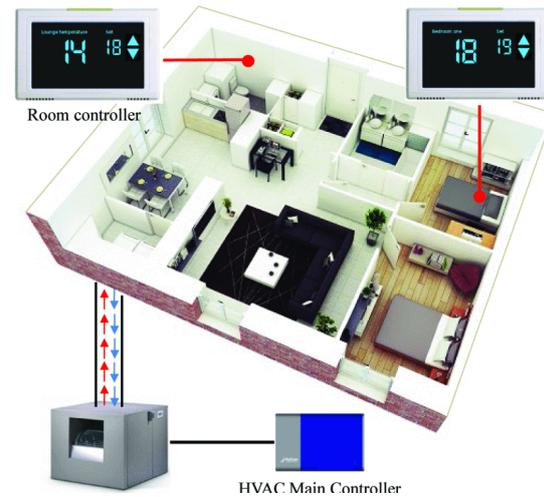

**Fig. 1:** Sub-systems of a typical building HVAC installation.

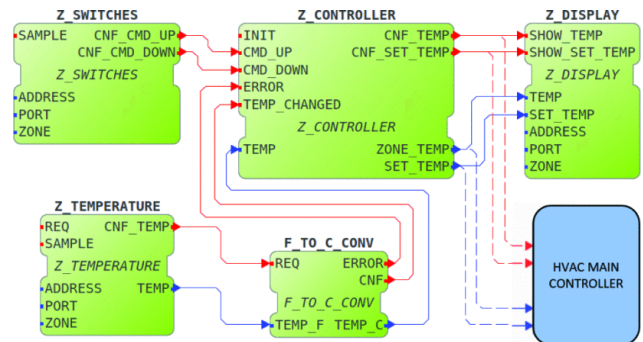

**Fig. 2:** Room controller built with IEC 61499 function blocks.

*Outputs* are used to pass information out of Z_CONTROLLER and Z_TEMPERATURE to other FBs.

Distributed, event-driven behavior such as this is typical of ICPS constructed from FBs. IEC 61499 facilitates the development of compact, modular functional units that encourage reusability. However, this abstraction level still demands a consideration of both the computational needs of the task as well as the physical aspects of the sensors, actuators and the work that is to be performed in the environment. ICPS operate in worlds that are non-deterministic. Hence their actions have to take into consideration the time-constraints that the environment imposes upon them. It is there, at the intersection of the cyber and the physical boundaries of an ICPS, that the most challenging aspects of its design need to be addressed [12]. Leitao [3] comments that the novelty of ICPS lies not in the establishment of new technologies. Rather, they draw together existing approaches from the domains of industrial control, real-time systems, service-oriented computing and distributed processing.

ICPS faults are defined as any operation that leads to unacceptable behavior or degraded performance [13], [14]. Sensors can include complex electronic interfaces that exhibit both breakdowns and anomalous behavior [15]. The electromechanical actuators that move the vanes in air ducts can also jam when they ice-over. Damaged position sensors can report their alignment incorrectly in these cases. Software faults, which are often indistinguishable from hardware faults, can be introduced into FBs that were previously working correctly via updates that have not been fully tested.

## 3. MULTI-AGENT SYSTEMS AND AGENT BELIEFS

Multi-Agent Systems first attracted interest in the 1980s as a way of performing control and management tasks within complex dynamic environments [16]. Bratman introduced the Belief-Desire-Intention (BDI) paradigm [17]. Traditionally, a BDI agent maintains beliefs about its environment, other agents, and itself as well as desires that it wishes to satisfy and intentions to act towards the fulfillment of selected desires. In the GORITE (Goal ORIented TEams) Multi-Agent Framework [18], desires are represented explicitly as goals [16]. When an agent commits to the achievement of a goal (based on its current beliefs), that goal becomes an intention. Further-more, GORITE agents (and teams) can be members of other teams which have their own beliefs, desires and intentions.

Beliefs provide a model of the domain the agent operates in [19]. For fault diagnosis, this model encompasses both the design of the ICPS that the agents are examining as well as sufficient knowledge of the physical environment the ICPS is interacting with. The agents gather all their knowledge of the physical environment via the ICPS itself. For example, the temperatures which the agents use to determine if the HVAC is working properly are thosecaptured by the FB that reads the room sensor. In our FDE design, reference temperatures are not captured by the agents themselves from other separate sensors co-located in the environment.

**Definition 1** (Beliefs). *Every agent contains a set of beliefs* $B = \{b_1,\ldots,\}$ *such that each belief* $b \in B$ *is a tuple* $\langle \Delta, v \rangle$ *where*

- $\Delta$ *is a skill that the agent can use.*
- $v$ *is the veracity of the belief held by the agent about the skill. This may be true, false or undetermined.*

In the FDE, agents can have three types of beliefs - *interaction* beliefs, *system-under-diagnostics* (*SUD*) beliefs and *dynamic diagnostics* beliefs.

### A. Beliefs about abilities to interact with other agents

Agents are imbued with beliefs about the skills and tools they can wield to perform goals.

**Definition 2** (Interaction belief). *A belief* $b = \langle \Delta, v \rangle$ *is an interaction belief when* $\Delta$ *describes a pair* $(\mathcal{A}, \mathcal{S})$. $\mathcal{A}$ *represents an agent and the* $\mathcal{S}$ *is the signature of a method that can be used by that agent to interact with the FBA and other agents.*

While pursuing its current goal, an agent instance knows how to communicate with other named agents in its team using skills such as Jane.say(*text*). Agents share diagnostic data on asynchronous FDE back channels where all communication is buffered in personal queues. The agent skill Jane.hear() retrieves the last message sent from that agent. This belief structure also models IEC 61499-specific domain knowledge needed to interact with the FBA being diagnosed.

Fig. 3 shows Diagnostic Points (DPs) that a FDE agent can use to interact with an application FB to capture telemetry and trigger test values. When an agent performs a rewire() interaction, a specialized DP diagnostic FB instance is inserted into an event and data path to enable it to read values and pass events transparently through the FBA. The gateClose() command isolates this information path so that a trigger() command can inject test values which are then captured further down the path from other DPs using the read() command. While other belief sets are dynamic, the capabilities captured in this first core set are static, intrinsic skills since they are core domain-specific abilities of the agent. Hence, the veracity of interaction beliefs is set to *true*. As a future direction of this work, a more dynamic setting can enable these beliefs to change over time.

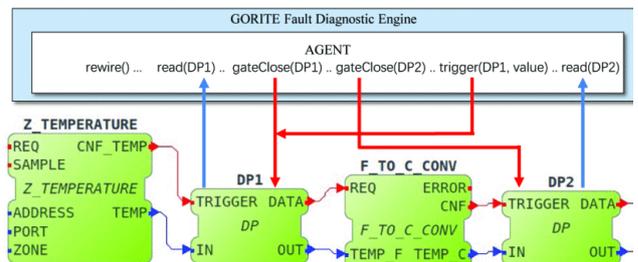

**Fig. 3:** Capturing and triggering Diagnostic Points.

### B. Beliefs about the FBA

Before the diagnostic goals are assigned, a second belief structure is made available to the agent that provides knowledge about the FBA and its structure.

**Definition 3** (System belief). *A belief* $b = \langle \Delta, v \rangle$, $ve$ *is called a system belief when is described by the triple* $\langle fb_1, trg, fb_2 \rangle$. $fb_1$ *and* $fb_2$ *are function block instances in the system under diagnostics, and trg represents the conditions* (*events and variable values*) *under which a transition can be triggered by the agent from* $fb_1$ *to* $fb_2$.

IEC 61499 Application Definition files are optimized to work with development IDEs such as 4diac [20]. However, their structure is hard for an agent to navigate since the parameters for each FB are stored in different parts of the XML-format file. Fig. 4 shows the agents belief structure about the FBA restructured as a Directed Graph. FBs are modeled as nodes and connections as directed edges. The name of an edge corresponds to the Output Event or Data Output on the FB the node describes. The internal data structure that holds this information is much easier for the agents to navigate when rewiring the FBA to insert DPs.

Fig. 5 shows the hierarchical organization of the information stored in each node about the individual FB. Agents are able to access the FB data during their diagnosis by referencing the single instance inside the FDE application belief data structure.

Each node of this belief structure provides the agent with detailed knowledge to allow it to reliably interact with each FB [4]. Agents are responsible for determining by themselves how to interact with a particular FB, verifying Input and Output data types so they can provide type-safe test values. The Directed Graph structure also allows agents to navigate fault paths autonomously, making decisions about the next possible fault location while they investigate. This belief structure is dynamic, providing the agents with a way to remember the result of diagnosing a FB as they traverse the FBA and update their beliefs about what is wrong.

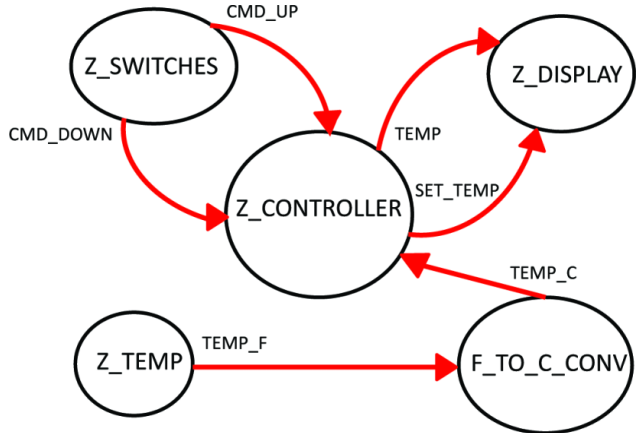

Fig. 4: The Function Block Application as a Directed Graph.

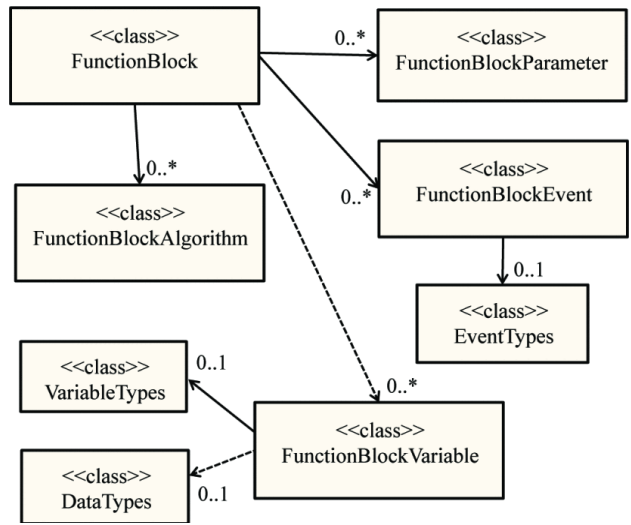

Fig. 5: Function Block information stored in a node.

### C. Beliefs about what is happening

As an agent performs a diagnosis, it forms new beliefs and modifies its existing beliefs.

**Definition 4** (Dynamic diagnostics belief). *A belief* $b=\langle \Delta, v \rangle$ *is a dynamic diagnostics belief if $\Delta$ is represented as a pair* $(fb_i, f_q)$ *where $fb_i$ is a function block instance of the system under diagnosis and $f_q$ is a valid fault code for $fb_i$, obtained from a set of fault codes F.*

In this belief structure, a belief about what is happening is a fuzzy logic opinion about a particular FB, sensor, actuator or algorithm in the FBA. The set $B = \{b_0, b_1, ..b_n\}$ captures the agent's beliefs about the FBA where $b_i$ represents either a single BFB or a network of BFBs connected as a discrete Composite Function Block (CFB). A BFB or CFB represents the smallest unit of functionality an agent can test and hence beliefs are atomic at this level.

During the design phase, a package of information was created that identifies the DPs available for each FB. These diagnostic packages contain sets of test values associated with data pathways through the FB that can be used during diagnosis to determine if the FB is performing correctly. The agent interprets these diagnostic packages while iterating each FB. A *diagnostic harness* is then created by inserting all the numbered DP instances into the FBA using rewire() commands. A belief is also established for each FB that has DPs that can be monitored.

Before the FDE instructs the agent to begin operating, the agent is provided with a set of goals that include monitoring for fault signatures and executing diagnostic plans. The agent is also provided with a definition of what constitutes normal behavior for this FBA. One example of normal behavior for the room controller shown in Fig. 2 is the appearance of temperature readings at 500ms intervals from Z_TEMPERATURE TEMP. These propagate through the controller to appear at Z_CONTROLLER ZONE_TEMP. Even when a temperature change has been requested, signaled by either a CMD_UP or CMD_DOWN event, the HVAC Industry ASHRAE Standard 55 specifies the optimal occupant thermal comfort rate of change should be a $\Delta t \pm 0.3°C$/min [21].

The agent pursues its GORITE Monitor goal, launching the FBA to run with its diagnostic harness enabled. All DP instances begin passing events and data through transparently to other FBs as well as back to the FDE agent. Initial evaluations with the room controller cycling at 100ms intervals showed no measurable degradation in performance with the diagnostic harness in-place. The agent continues pursuing its Monitor goal until one or more of its primary beliefs about normal operation are invalidated. Fig. 6 shows the rewired Room Controller with DPs inside the temperature sub-system. The agent establishes an initial belief for the Monitor goal such that:

$$b_0 = \langle fb_{Z\_TEMPERATURE}, v_{undetermined}, f0 \rangle$$

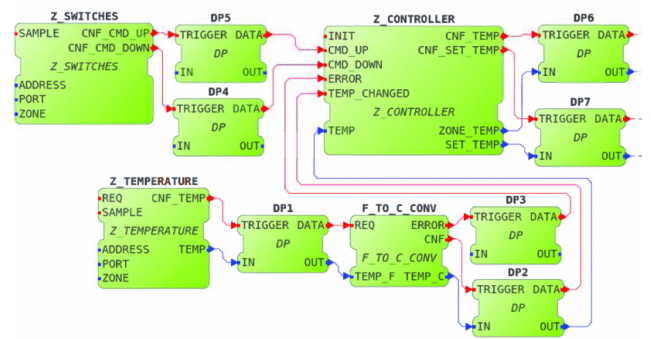

Fig. 6: Rewired Room Controller with Diagnostic Points.

If the result of the agent using the primary monitoring plan to look for normal behavior determines that the FBA is operating within tolerance then $v_{undetermined} \rightarrow v_{true}$ as the agent reinforces this primary belief. Other primary beliefs monitor the response from the room controller and the HVAC Main Controller to a CMD_UP or CMD_DOWN being issued by Z_SWITCHES. For an agent, the pursuit of its Monitor goal is the repeated re-evaluation of each of its beliefs by running the prescribed tests at defined intervals.

Invalidating any one of these beliefs in *B* causes the agent to signal its GORITE Team that it is abandoning its primary Monitoring goal in favor of pursuing its Diagnostic goal. The agent adopts a divide-and-conquer strategy for diagnosing faults in the temperature sensor sub-system of the room controller. The agent first issues gateClose() commands to some of the DP instances within the temperature sub-system. This disconnects Z_TEMPERATURE which is responsible for delivering the temperature sensor readings. Z_SWITCHES is isolated by gating DP4 and DP5 and updates to the HVAC Main Controller are blocked by gating DP6 and DP7. The agent then uses the set of tests for F_TO_C_CONV to exercise its events, inputs and outputs. A range of nominal Fahrenheit temperatures are injected via DP1 and captured as Celcius values at DP2. Out-of-range values such as absolute zero ($-459.67^oF$) should trigger the ERROR event, captured by DP3. If all test values are converted and captured correctly, the agent updates the *v* of the $b_{F\_TO\_C\_CONV}$ to true. The agent continues down the data path checking each subsequent FB and updating its beliefs until all the FBs have been tested. This process caters for the possibility of multiple fault candidates.

In subsequent Analyse and Report goals, the agent proposes a diagnosis after iterating each belief to examine its veracity. We simulated a number of fault scenarios to determine how the agent would respond in each case. A software update failure was simulated by modifying the Fahrenheit to Celsius algorithm in F_TO_C_CONV to output random readings when a test value was between $70°F$ and $80°F$. The randomizer used to simulate the fault was tuned so that it did not always generate errors in every test run. This caused the agent to miss this intermittent behavior during some diagnostic sessions. However, the FDE correlates multiple diagnostic sessions to capture a more comprehensive diagnosis and this discrepancy was reported to be an intermittent rather than a hard fault.

Since sensors cannot usually be tested directly, the agent established an initial belief for the temperature sensor Z_TEMPERATURE as $v_{undetermined}$. This returned a hypothesis that the sensor may be faulty if all other FBs pass their diagnostic tests. In all situations, any belief that cannot be verified is reported as a possible fault with an implied lower probability.

We examined a number of criteria while evaluating the performance of our prototype FDE. GORITE is designed to host multiple agents, efficiently co-ordinating goal sharing and interaction while allowing the agents sufficient headroom to perform their tasks. A GORITE sequential BDI Goal execution mode was implemented for the FDE, allowing the agent to step from Monitor goals into Diagnose, Analyse, and Report goals. The performance of the GORITE framework with the FBA executing on 100ms cycles required the insertion of loops and multi-second delays during goal execution to give the agent sufficient time to capture all telemetry. With multiple agents sharing tasks, this suggests that the framework still has sufficient headroom left to allow the FDE to scale well. The DP probe FBs are custom designed in C++ to be highly-efficient, managing their own communications channels with their agent using internal TCP/IP clients implemented in each instance.

## 4. CONCLUSIONS AND FUTURE WORK

This paper demonstrates how relatively simple diagnostic test scenarios, created during the design and creation of each FB, provide a way to employ agents to perform sophisticated fault monitoring and diagnosis. Since many fault scenarios are elusive and only appear under certain conditions, deploying agents for long-term monitoring during commissioning or burn-in trials is a promising approach. By ensuring that the DP FBs remain lightweight probes rather than agents in their own right helps to mitigate the effects of introducing additional FBs into the application. This allows more realist evaluations of performance issues such as timing.

Future directions for this work include plug-ins for creating FB diagnostic packages directly inside IEC 61499 development systems such as 4diac. These would assist engineers while they are building FBs. The ability to deploy an agent to verify FBs iteratively addresses some of the shortcomings of Unit Testing and Test-Driven Design in current design tools. Analyzing the Execution Control Chart (ECC) that drives state transitions in a FB is one possible way of identifying DPs automatically when creating diagnostic packages.

Collaboration between teams of distributed agents also offers a way of addressing the challenges of modeling in these environments. The dynamic models the agents create by themselves from IEC 61499 design artifacts highlights the value of the semi-autonomous design-time support agents could provide.